\documentclass[12pt]{iopart}
\usepackage{graphicx}
\begin{document}

\title{Reflection of OH molecules from magnetic mirrors}
\author{Markus Mets\"al\"a, Joop J. Gilijamse, Steven Hoekstra,
Sebastiaan Y. T. van de Meerakker, and Gerard Meijer}

\address{Fritz-Haber-Institut der Max-Planck-Gesellschaft, Faradayweg 4-6, 14195 Berlin, Germany}

\begin{abstract}
We have reflected a Stark-decelerated beam of OH molecules under
normal incidence from mirrors consisting of permanent magnets. Two
different types of magnetic mirrors have been demonstrated. A
long-range flat mirror made from a large disc magnet has been used
to spatially focus the reflected beam in the longitudinal
direction ("bunching"). A short-range curved mirror composed of an
array of small cube magnets allows for transverse focusing of the
reflected beam.
\end{abstract}

\pacs{32.60.+i, 37.20.+j, 37.10.Mn, 41.20.Gz}

\section{Introduction}

The use of inhomogeneous magnetic fields to influence the
trajectories of atoms and molecules in free flight has played an
important role in the development of the field of atomic and
molecular beams and has contributed enormously to its
success~\cite{Scoles}. The manipulation of beams of atoms and
molecules with external magnetic fields as it has been used in the
past almost exclusively involved the transverse motion of the
particles~\cite{Gerlach,Rabi}. The reason for this is obvious: the
magnetic field gradients that can be realized in the laboratory
are sufficient to influence the transverse velocity components of
the molecules in the beam, as these are centered around zero
velocity. However, the forward velocity of molecules in
conventional beams is centered around a large value. Therefore, to
be able to influence the longitudinal motion of the molecules slow
beams are required.

During the last years a variety of methods have been demonstrated
to produce beams of slow molecules. Several of these methods use
seeded, pulsed molecular beams as a starting point. In such beams
only a limited number of ro-vibrational levels is populated, with
a high phase-space density. To reduce the speed of the molecules
in the beam, the interaction of the molecules with time-varying
electromagnetic fields can be exploited. Thus far, this has been
demonstrated for molecules using electric
(\cite{Bethlem99,Heiner06} and references therein) and optical
fields~\cite{Barker}, and for atoms using magnetic
fields~\cite{Vanhaecke, Narevicius}. Mechanical means to produce
beams of slow molecules have also been demonstrated, e.g.~the
back-spinning nozzle~\cite{Herschbach}. Closely related to this
are methods that utilize reactive~\cite{Loesch} or rotationally
inelastic~\cite{Chandler} collisions in counter-propagating or
crossed beams to produce slow molecules. Alternatively, rather
than starting from a pulsed supersonic beam, the low-velocity tail
of the thermal Maxwellian velocity distribution in an effusive
molecular beam can be filtered out~\cite{Rempe}. Via all of these
approaches, beams of molecules with kinetic energies on the order
of 1~cm$^{-1}$ or less can be produced. This is sufficiently low
to allow for normal incidence reflection from magnetic mirrors.

In the field of cold atoms, permanent magnets have found many
uses, and samples of cold atoms have been reflected, guided and
trapped by their potentials~\cite{Hinds}. As samples of cold atoms
are routinely produced with temperatures in the $\mu$K range, the
requirements on the magnetic field-strengths are rather relaxed.
Periodically magnetized recording media, e.g., video or audio
tape, have been used to construct curved mirrors
for atoms. In the experiments presented here, we have used beams
of ground-state OH molecules, slowed down using a Stark
decelerator. These decelerated beams still have a kinetic energy
corresponding to a temperature of a few hundred mK, and
significantly higher field-strengths are therefore required. We
have used rare-earth magnets that can possess a remanence of
higher than 1~T. The small size and the low cost of these magnets
makes them ideally suited to create different types of magnetic
mirrors for molecules.

\section{Experimental setup}

\begin{figure}[t!]
\begin{center}
\includegraphics{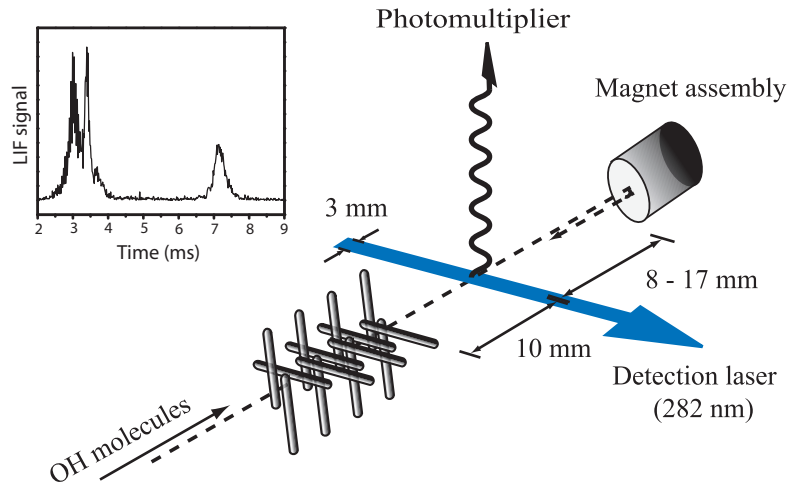}
\caption{A schematic view of the detection zone in the experiment.
For clarity, only the last seven electrode pairs of the
decelerator are shown. The dashed line indicates the path of the
molecular beam. The graph in the upper left corner shows a
time-of-flight profile for a beam of ground-state OH molecules
decelerated from 390~m/s to 15~m/s. The undecelerated molecules
arrive in the detection zone about 3~ms after their production in
the source region, whereas the decelerated ones take some 4~ms
longer.} \label{Metsala-Figure1}
\end{center}
\end{figure}

In Fig.~\ref{Metsala-Figure1} a scheme of the essential part of
the experimental setup is shown. A detailed description of the
molecular beam machine, and particularly of the deceleration of a
beam of OH molecules, is given
elsewhere~\cite{Meerakker05,Meerakker06}. In the experiments
reported here, a pulsed beam of OH molecules is decelerated from
390~m/s to a velocity in the 10-20~m/s range using a 119~cm long
Stark decelerator consisting of 108 electric field stages; in
Fig.~\ref{Metsala-Figure1} only the last six stages of the Stark
decelerator are shown. About 10~mm behind the decelerator, the OH
molecules pass through the detection zone. Here, laser induced
fluorescence (LIF) excitation is performed by 282~nm laser light
from a pulsed dye laser (5~ns pulse duration, bandwidth of about
0.1~cm$^{-1}$). The $Q_1(1)$ transition of the $A^2\Sigma^+,v=1
\leftarrow X^2\Pi_{3/2},v=0$ band is induced and the resulting
off-resonant fluorescence to the $X^2\Pi, v=1$ state is recorded
by a photomultiplier tube. About 2~mJ of laser pulse energy in a
3~mm diameter beam is used, which is sufficient to saturate the
transition. The experiment runs at a 10~Hz repetition frequency,
and a typical measurement is averaged over 10-1000 shots for
improved statistics.

A typical time-of-flight measurement for a beam of OH molecules
passing through the Stark decelerator beam machine is shown in the
upper left corner of Fig.~\ref{Metsala-Figure1}. The undecelerated
part of the OH beam arrives in the detection region about 3~ms
after its production in the source chamber. Only OH
($X^2\Pi_{3/2}, v=0, J=3/2$) molecules in the low-field seeking
component with the largest Stark shift, e.g. the $M_J\Omega=-9/4$
component, are decelerated. In this particular measurement, these
molecules are decelerated to a final velocity of 15~m/s and they
arrive in the detection region after 7~ms. The decelerated packet
contains approximately $10^5$-$10^6$ OH molecules, has a spatial
extent of about 3~mm along the molecular beam axis and is about
4x4~mm$^2$ in the transverse direction at the exit of the
decelerator. The full width at half maximum (FWHM) of the velocity
spread in the forward direction is about 7~m/s, corresponding to a
longitudinal temperature of 10~mK. In the transverse direction,
the FWHM velocity spread is slightly less, about 5~m/s. Only this
decelerated packet of OH molecules is relevant for the magnetic
reflection experiments presented here.

Shortly behind the detection region, two different magnet
assemblies can be mounted on a vacuum feedthrough translator that
allows to change the distance of the magnet surface to the
detection zone from 0 to 30~mm. Ground state OH molecules are
ideally suited for Stark deceleration and subsequent magnetic
reflection because they exhibit both a strong Stark shift and a
strong Zeeman shift; Stark deceleration followed by
magnetoelectrostatic trapping has first been demonstrated with OH
molecules as well~\cite{Sawyer}. The interaction potential of the
ground state ($X^2\Pi_{3/2}, J=3/2$) OH molecules with the
magnetic field is given by the first-order Zeeman effect. Half of
the decelerated OH molecules are in the magnetically high-field
seeking states and will be attracted to the magnet, i.e., they
will crash into the magnet and be lost for the further experiment.
The magnetically low-field seeking OH molecules will experience a
repulsive force, and after reflection they can pass through the
detection region once more. For the $M=+3/2$ state the Zeeman
shift is equal to 0.56~cm$^{-1}$/T. For a magnetic field of 0.5~T
this implies that OH molecules with velocities up to a maximum of
about 20~m/s can be retro-reflected.

\begin{figure}[t!]
\begin{center}
\includegraphics{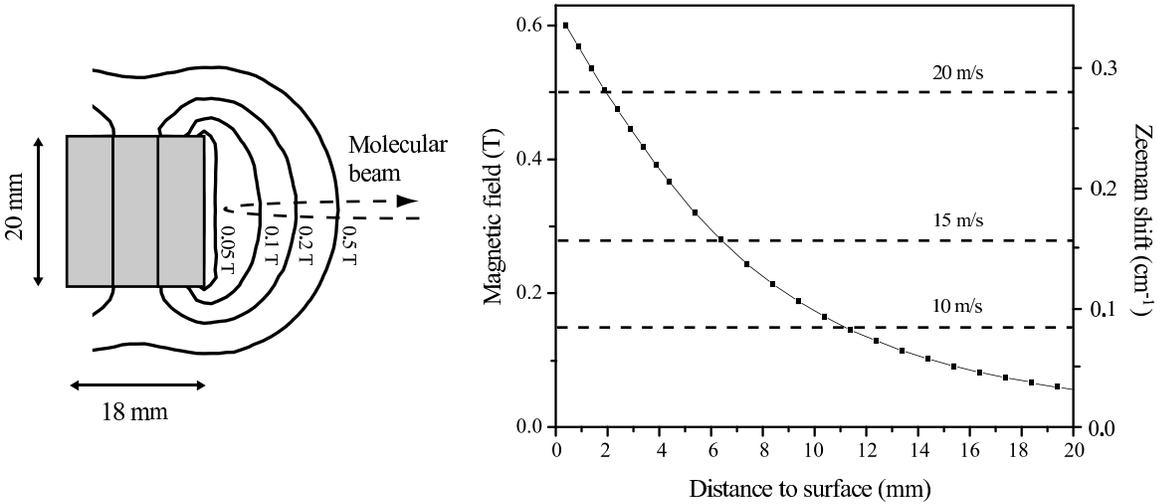}
\caption{Left panel: Schematic picture of the disc magnet
including calculated iso-magnetic-field lines at 0.05, 0.1, 0.2,
and 0.5~T. Right panel: The measured magnetic field strength and
the corresponding Zeeman shift for decelerated OH ($M=+3/2$)
molecules as a function of distance from the surface of the disc
magnet. The dashed horizontal lines indicate the field strengths
at which OH molecules with velocities of 10, 15 and 20~m/s will be
reflected.} \label{Metsala-Figure2}
\end{center}
\end{figure}

The first magnetic mirror consists of a stack of three disc-shaped
nickel coated NdFeB magnets, each with a diameter of 20~mm and a
thickness of 6~mm, and is referred to hereafter as the "disc
magnet". The remanence of the material used for the magnets is
1.5~T according to the supplier. The measured magnetic field and
the corresponding Zeeman shift for the $M=+3/2$ state along the
centerline of the disc magnet as a function of distance from the
magnet surface is given in Fig.~\ref{Metsala-Figure2}. The dependence of
the magnet field strength on distance from the surface is rather
similar for axes that are parallel to and up to a few mm away from
the centerline; the magnetic field strength only starts to
drop off significantly near the edges of the magnet. This mirror
therefore acts more or less as a flat mirror. It is observed that
the magnetic field has a fairly long range, and that even at
distances of a few centimeters from the magnet surface there is
still an appreciable magnetic field present. The magnetic field at
which OH molecules with a velocity of 10, 15 and 20~m/s will be
reflected is indicated by the horizontal dashed lines in
Fig.~\ref{Metsala-Figure2}. It can be seen that OH molecules with a
velocity of 20~m/s will penetrate about 9~mm further into the
magnetic field than molecules that move with only half that
velocity. This leads to bunching, i.e. a longitudinal spatial
focusing, of the reflected molecules.

\begin{figure}[t!]
\begin{center}
\includegraphics{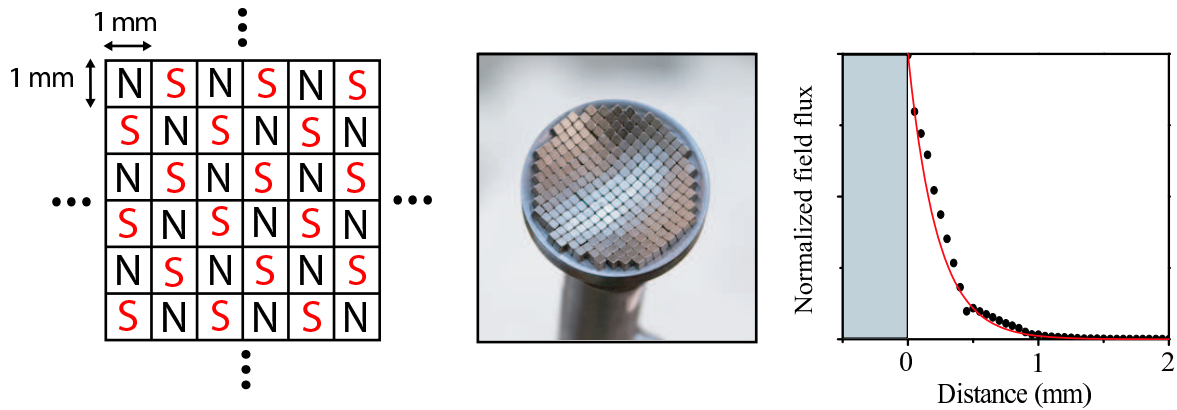}
\caption{Left panel: schematic view of the two-dimensional
magnetic array. N and S indicate north and south poles of the
individual magnets. Center panel: Photograph of the actual array.
Right panel: Magnetic field strength as a function of distance
from the surface of the array. The solid curve is the analytical
expression (Eq.~\ref{decay}) whereas the dots are the result of a
finite element calculation.} \label{Metsala-Figure3}
\end{center}
\end{figure}

The second magnetic mirror is a two-dimensional array of 241
nickel coated 1x1x1~mm$^3$ cube NdFeB magnets arranged on a
concave round iron substrate with a diameter of 20~mm and a radius
of curvature of 20~mm. A photograph of this magnetic mirror,
hereafter referred to as the "magnet array", is shown in the
center panel of Fig.~\ref{Metsala-Figure3}. The remanence of the
material used for the magnets is 1.2~T according to the supplier.
The cube magnets are arranged in a checkerboard fashion, with
alternating north and south poles in two dimensions, as is
schematically depicted in the left panel of
Fig.~\ref{Metsala-Figure3}.

If the magnetization is varying sinusoidally with a period of $a$
in the {\it x-y} plane, then the decay of the magnetic field in {\it z}
direction can be derived similarly as is done for the one dimensional
case~\cite{Hinds,Opat}. For the two-dimensional case the field strength
$B$ will decay as
\begin{equation}
B=B_0e^{-\sqrt{2}kz}
\label{decay}
\end{equation}
where $B_0$ is the field strength at the surface of the array and
$k=2\pi/a$. For our case of $a=2$~mm we have a $1/e$ decay
distance of slightly more than 0.2~mm. We neglect here the higher
harmonics of the alternating magnetization and end effects due to
the finite number of magnets~\cite{Sidorov}. To validate this result
we performed a finite element calculation, the result of which is
shown in the right panel of Fig.~\ref{Metsala-Figure3} (dots)
together with the predicted field strength from Eq.~\ref{decay}
(solid curve). As the magnetic field of this mirror extends only
over a short range, the molecules will all be reflected at
more or less the same distance from the surface, irrespective of
their velocity. It is evident, however, that this magnet array
forms a transversely focusing mirror with a
focal length of 10~mm; after the addition of the magnet array the
curvature of the iron substrate is not exactly preserved and the
focal distance is actually slightly shorter than this.

An analogous electrostatic mirror, constructed with a
one-dimensional array of thin electrodes, has been used to
retro-reflect a slow beam of ammonia molecules~\cite{Schultz}. A
one-dimensional array of small permanent magnets has been used for
reflection of cold cesium atoms dropped from an optical
molasses~\cite{Sidorov}. By adding a magnetic bias field to the
magnet array, as used here, a two-dimensional array of small traps
with large magnetic field gradients can be
created~\cite{Sinclair,Gerritsma}.

\section{Results and discussion}

\begin{figure}[t!]
\begin{center}
\includegraphics{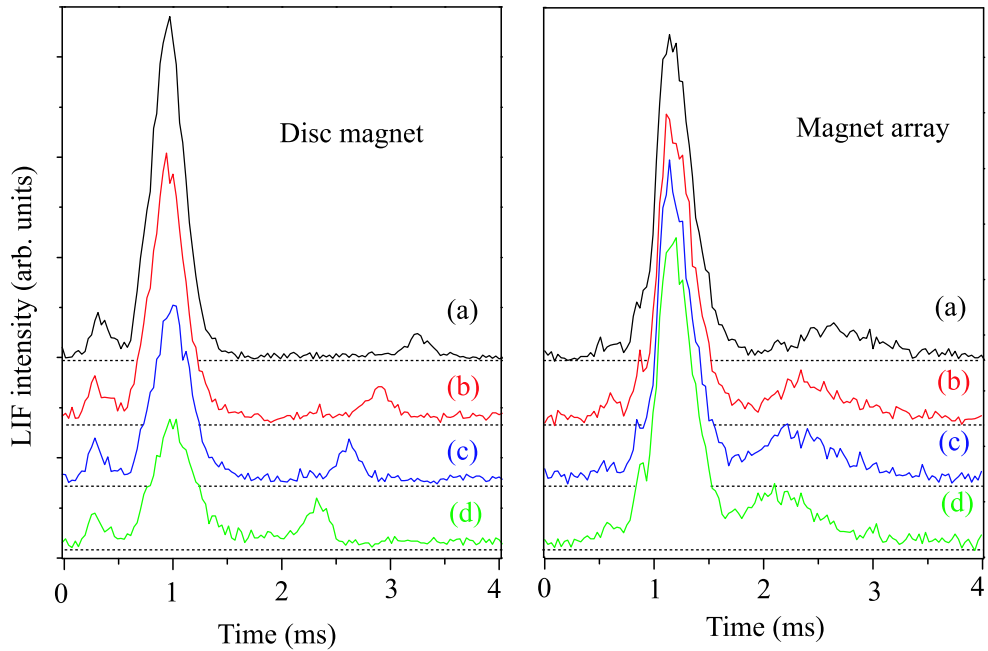}
\caption{Measured time-of-flight profiles for decelerated beams of OH
molecules reflected from two different magnetic mirrors. Left panel:
OH molecules with a mean velocity of 19~m/s are reflected from the
disc magnet. The distance of the surface of the magnet to the detection
zone is (a) 17~mm, (b) 13~mm, (c) 10~mm and (d) 8~mm. Right panel:
OH molecules with a mean velocity of 15~m/s are reflected from the
magnet array. The distance of the surface of the magnet to the detection
zone is (a) 13~mm, (b) 11~mm, (c) 10~mm and (d) 8~mm.}
\label{Metsala-Figure4}
\end{center}
\end{figure}

The left panel of Fig.~\ref{Metsala-Figure4} shows time-of-flight
(TOF) profiles obtained by reflecting OH molecules from the disc
magnet. The data shown are an average over 350 measurements. For
convenience, we define the time that the electric fields of the
decelerator are switched off, e.g. when the decelerated molecules
are near the exit of the decelerator, as $t=0$. The main peak at
around 1~ms results from the decelerated packet of OH molecules
that passes with a mean velocity of 19~m/s through the detection
zone on its way to the magnetic mirror. The reflected packet of
molecules returns in the detection zone up to several milliseconds
later. The exact arrival time obviously depends on the distance of
the magnetic mirror from the detection zone which is varied from
17 via 13 and 10 to 8~mm for curves (a)--(d), respectively. The
small peak at around 0.3~ms results from OH molecules that are
considerably faster than the decelerated packet, and that will
therefore not be reflected. One can see in
Fig.~\ref{Metsala-Figure4} that the height of the main peak at
around 1~ms is lower when the magnet surface is placed closer to
the detection region. This is due to the less efficient detection
of the OH molecules as a consequence of the Zeeman broadening of
the detection transition, caused by the long-range field of the
disc magnet. In addition, the background signal due to laser light
scattered from the disc magnet increases when the magnet is
brought closer to the detection zone.

Normally, one would expect the peak of the reflected beam in the
TOF profile to be considerable broader than that of the incoming
beam, given the longer flight time in combination with the
relatively large longitudinal velocity spread in the decelerated
beam. The TOF profiles clearly show, however, that the reflected
peak is even narrower than the incoming one. This is a direct
demonstration of longitudinal spatial focusing of the reflected
beam, caused by the long-range field of the disc magnet. The
faster molecules penetrate deeper into the magnetic field than the
slow ones, and therefore have to travel
a longer distance. When the distance of the surface of the disc
magnet to the detection region is about 10~mm (c), this results in
a catching up of the slow and the fast molecules in the detection
region, i.e. in a longitudinal spatial focus in the detection
region and in the corresponding narrowest peak in the TOF
distribution. Longitudinal spatial focusing, both in real space
(bunching) and in velocity space (longitudinal cooling), of a
molecular beam using inhomogeneous electric fields has been
discussed and demonstrated before~\cite{Crompvoets}.

The right panel of Fig.~\ref{Metsala-Figure4} shows TOF profiles
obtained by reflecting OH molecules from the magnet array. The
data shown are an average over 660 measurements. Because the
strength of the magnetic field is lower than in the case of the
disc magnet, the OH molecules are decelerated to a mean velocity
of 15~m/s for these experiments. It is seen in
Fig.~\ref{Metsala-Figure4} that the signal due to the incoming
slow OH beam remains constant even when the surface of the magnet
array is rather close to the detection zone; only the stray light
level increases due to light scattering from the magnet array. As
expected, there is no evidence of longitudinal spatial focusing of
the reflected beam, and the peak of the reflected beam in the TOF
distribution is considerably broader than that of the incoming
beam. It is interesting to note that for the same distance between
the surface of the magnetic mirror and the detection zone, the
reflected molecules arrive earlier when reflected from the magnet
array than from the disc magnet. This might seem
counter-intuitive, as the molecules in this experiment are not
only slower to begin with but also approach the surface of the
magnet array closer than that of the disc magnet, and therefore
have to travel a longer distance. However, with the disc magnet
the molecules are gradually decelerated as they approach the
magnet and they are accelerated again on their way to the
detection region. With the magnet array, on the other hand, the
molecules keep a constant speed throughout, with an abrupt change
of sign of the velocity vector close to the surface, and are
therefore actually earlier back in the detection region. We do see
an effect of the transverse spatial focusing of the magnet array.
For a distance of the surface of the magnet to the detection
region of 8 or 10~mm, the ratio of the area of the reflected peak
to the initial peak is about a factor two larger for the magnet
array than for the disc magnet. This indicates that the molecules
are detected more efficiently when reflected from the magnet array
due to the transverse focusing of the packet in the detection
region.

\begin{figure}[t!]
\begin{center}
\includegraphics{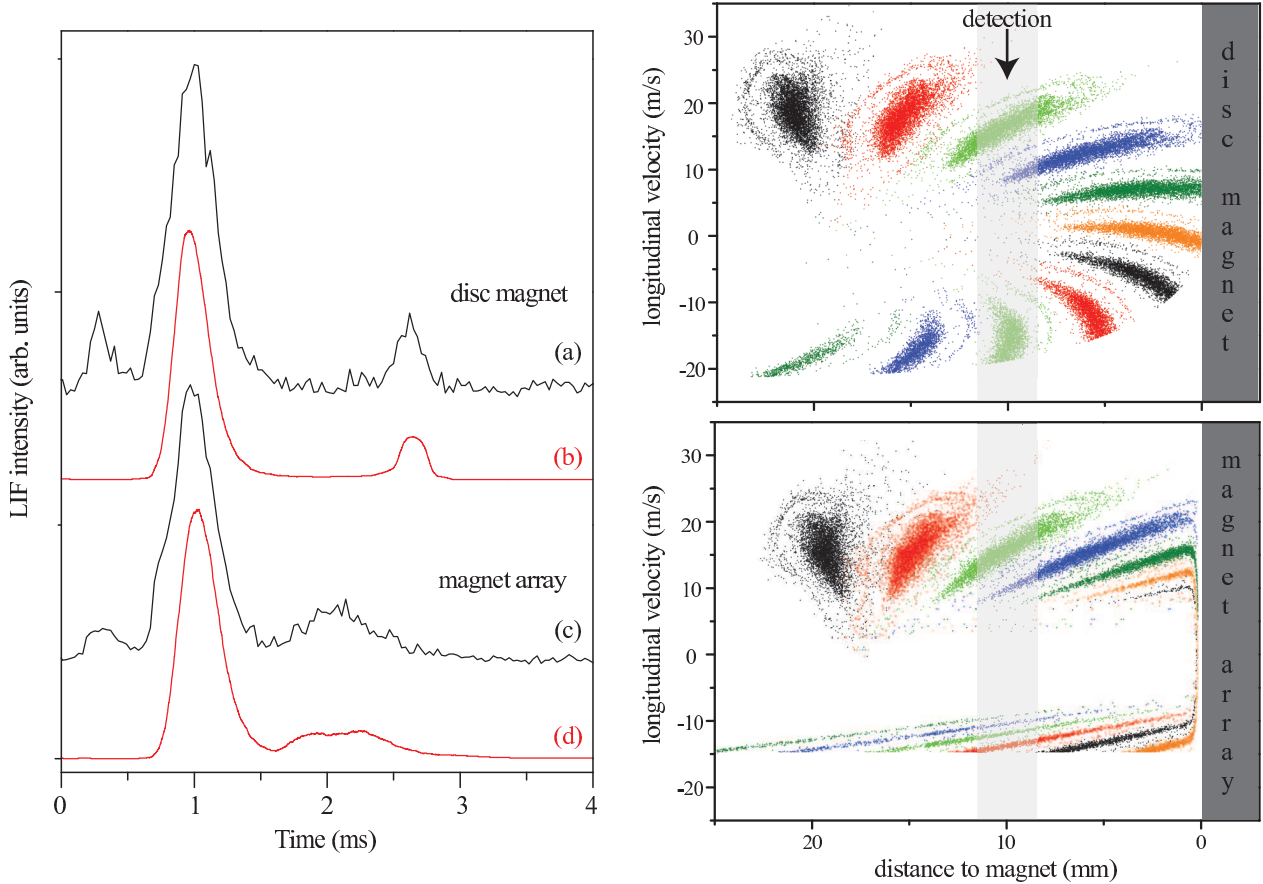}
\caption{Left panel: measured (a,c) and simulated (b,d) TOF
profiles for decelerated OH molecules reflected from the two
different magnetic mirrors. An OH beam with a mean velocity of
19~m/s (15~m/s) is reflected from the disc magnet (magnet array).
The distance of the surface of the magnetic mirror to the
detection zone is 10~mm. Right panel: snap-shots of the
corresponding simulated longitudinal phase-space distributions at
0.25~ms time intervals. The position of the detection laser is
indicated in grey.} \label{Metsala-Figure5}
\end{center}
\end{figure}

To further illustrate the effect of longitudinal spatial focusing
for the disc magnet, as well as the lack thereof for the magnet
array, we have done three-dimensional trajectory calculations to
simulate the reflection experiments. Fig.~\ref{Metsala-Figure5} shows
measured TOF profiles for both magnetic mirrors, together with the
corresponding simulated profiles. In the simulation for the disc
magnet the measured magnetic field as given in Fig.~\ref{Metsala-Figure2}
is taken, whereas for the magnet array the analytical expression
of Eq.~\ref{decay} with $B_0=0.3$~T is used. The shape and time
of arrival of the reflected peak is reproduced well in the
simulated TOF profiles for both magnetic mirrors. On the right
hand side of Fig.~\ref{Metsala-Figure5} snap-shots of the longitudinal
phase-space distributions of the OH molecules are shown at 0.25~ms
time intervals during the reflection process. As the OH molecules
approach the surface of the disc magnet, the fast molecules are
decelerated more than the slow ones and as a result the velocity
distribution narrows down while the position distribution broadens.
After turning around the opposite happens and the
spatial distribution comes to a focus in the detection region. In
the reflection process, the initial phase-space distribution of
the decelerated packet is more or less reconstructed at the
detection zone, but spatially compressed (and therefore slightly
extended in velocity space) and rotated by 180$^{\circ}$. The
clipping of the phase-space distributions at velocities of about
$-21$~m/s results from the finite height of the magnetic field at
the surface; faster molecules have simply not been reflected but
have crashed into the surface. The phase-space distributions for the OH
molecules that are reflected from the magnet array clearly show
that, after an abrupt reversal of the velocity close to the surface,
the reflected packet continues to spread out and that no
bunching occurs.

\section{Conclusions}

We have demonstrated reflection of slow beams of OH molecules from two
different types of magnetic mirrors, composed of permanent magnets.
A long-range flat mirror is used for longitudinal spatial focusing of
the reflected molecules whereas a short-range curved mirror is used for
transverse focusing.

Permanent magnets with typical magnetic field strengths of 1~T can
be custom-made in a wide variety of geometries. They are compact
and relatively cheap and can be a viable alternative to
electromagnets in the manipulation and control of cold molecules.
Apart from magnetic mirrors, magnetic guides and magnetic traps
can be made from these permanent magnets as well; the group of Jun
Ye (JILA, Boulder, CO, USA) has recently demonstrated magnetic
trapping of OH molecules using an arrangement of permanent
magnets~\cite{Ye-PC}. Even though permanent magnets do not allow
for rapid tuning of the magnetic field strength, often the
strength of the interaction of molecules with the magnetic field
can be tuned by (laser) preparation of the molecules in an
appropriate quantum state. For molecules in a $^2\Pi$ electronic
state like OH, the magnetic field interaction can even effectively
be switched off by transferring the molecules from the
$^2\Pi_{3/2}$ to the $^2\Pi_{1/2}$ electronic state. When the
transfer of molecules is performed via laser excitation followed
by spontaneous fluorescence, this scheme allows for the
accumulation of Stark decelerated molecules in an electrostatic
trap that is superimposed with the magnetic reflection field,
i.e., a more general version of the previously proposed
accumulation scheme~\cite{reloading}.

\section*{Acknowledgments}
We acknowledge the technical support from Manfred Erdmann. M.M. is
grateful to the Academy of Finland for financial support.

\end{document}